\documentclass[traditabstract]{aa}  

\usepackage[applemac]{inputenc}
\usepackage[english]{babel}
\usepackage{graphicx}
\usepackage{txfonts}
\usepackage{ulem}
\usepackage{color}
\usepackage{longtable}
\usepackage{multirow}

\begin{document}

\title{Local stability of a gravitating filament: a dispersion relation}

\titlerunning{Local stability of a gravitating filament}
\authorrunning{J.~Freundlich, C. Jog \& F. Combes}

\author{J.~Freundlich\inst{1}\fnmsep\thanks{\email{jonathan.freundlich@obspm.fr}}
          \and
          C. J.~Jog\inst{2}
          \and
          F.~Combes\inst{1}
         }


\institute{
	LERMA, Observatoire de Paris, CNRS, 61 av. de l'Observatoire, 75014 Paris, France
	\and
	Department of Physics, Indian Institute of Science, Bangalore 560012, India
	}

\date{Received ; accepted}

\abstract{Filamentary structures are ubiquitous in astrophysics and are observed at various scales. On a cosmological scale, matter is usually distributed along filaments, and filaments are also typical features of the interstellar medium. Within a cosmic filament, matter can contract and form galaxies, whereas an interstellar gas filament can clump into a series of bead-like structures that can then turn into stars. 
To investigate the growth of such instabilities, we derive a local dispersion relation for an idealized self-gravitating filament and study some of its properties. 
Our idealized picture consists of an infinite self-gravitating and rotating cylinder with pressure and density related by a polytropic equation of state. We assume no specific density distribution, treat matter as a fluid, and use hydrodynamics to derive the linearized equations that govern the local perturbations. 
We obtain a dispersion relation for axisymmetric perturbations and study its properties in the ($k_R$, $k_z$) phase space, where $k_R$ and $k_z$ are the radial and longitudinal wavenumbers, respectively. While the boundary between the stable and unstable regimes is symmetrical in $k_R$ and $k_z$ and analogous to the Jeans criterion, the most unstable mode displays an asymmetry that could constrain the shape of the structures that form within the filament. 
Here the results are applied to a fiducial interstellar filament, but could be extended for other astrophysical systems, such as cosmological filaments and tidal tails. 
}

\keywords{gravitation --
		hydrodynamics --
		instabilities --
		cosmology: large-scale structure of the Universe --
		ISM: structure
		}
		
\maketitle

\section{Introduction}

\subsection{Context}

Filamentary structures are ubiquitous on cosmological scales, forming a cosmic web that connects galaxies to one another (e.g., \cite{bond, springel, fumagalli}) and provides a gas reservoir from which galaxies grow and accrete (e.g., Kere\v{s} et al. 2005, Dekel et al. 2009). Sheets and filaments arise naturally as a consequence of asymmetrical gravitational collapse, because ellipsoidal maxima collapse at different speeds along the three unequal axes (\cite{lin}). 
The baryon distribution is expected to follow the dark matter distribution on large scales, but to differ on smaller scales. Simulations by Harford et al. (2008) have shown that intergalactic gas tends to concentrate towards the centers of filaments and sheets, with quasi-spherical dark matter structures at the intersections of the filaments. The inner core of many of these cosmic filaments is predominantly composed of gas, justifying models that treat them as self-gravitating, isothermal, or barotropic cylinders in hydrostatic equilibrium. 
Simulations show that when the cooling time is short enough, the inner part of the filament collapses to form dense cold cores that contain a substantial fraction of the gas (\cite{gray}).

In the interstellar medium, observations show filamentary structures on much smaller scales (\cite{andre}, \cite{arzoumanian}, \cite{jackson}, \cite{kirk}, \cite{palmeirim}). Motivated by \textit{Herschel} observations of star-forming environments, Andr\'e et al. (2010, 2013) suggest a scenario in which the formation of turbulence-driven filaments in the interstellar medium represents the first step towards core and star formation. The densest filaments would then fragment into prestellar cores owing to gravitational instability. 
These filaments have typical widths of about 0.1 parsec, which could be explained by the combination of accretion-driven turbulence and dissipation by ion-neutral friction (\cite{arzoumanian}, \cite{hennebelleandre}). 
Simulations also reveal filamentary features, arising either from turbulence (Padoan et al. 2001) or from intermediate stages of gravitational collapse (Gomez et al. 2013). 

The standard Jeans instability describes the collapse of a spherical gas cloud when the inner pressure is not strong enough to support the self-gravitating gas. Small perturbations are amplified when their scales are larger than a specific length, the Jeans length (e.g., \cite{binney}). The cylindrical case is less straightforward and has not been fully investigated yet. The derivation of a dispersion relation for the perturbations arising in a cylindrical filament would enable a better understanding of the behavior of such perturbations. This is the motivation of the present paper.

\subsection{Previous work}

Interest in the dynamics of cylindrically symmetric systems has revived in recent years, motivated by the filamentary structures revealed in observations and simulations. But early studies have taken a purely theoretical point of view. 
Chandrasekhar \& Fermi (1953) studied the dynamics of a homogeneous and incompressible, infinite, self-gravitating cylinder with a constant axial magnetic field, while Ostriker (1964b) generalized their calculations to a homogeneous compressible cylinder. 
Simon (1963) also carried out a perturbation analysis for such a homogeneous compressible cylinder, and Stod\'olkiewicz (1963) considered an isothermal compressible cylinder embedded in different types of magnetic fields and derived criteria for the critical length of the perturbations. 

In his seminal 1964 paper, Toomre carried out a perturbation analysis for a rotating disk and obtained a dispersion relation assuming local perturbations. 
The calculation applies to galactic disks, but not to elongated filaments. 
Mikhailovskii \& Fridman (1972, 1973) and Fridman \& Polyachenko (1984) later described the instability of a homogeneous gravitating cylinder of finite radius and infinite length, using a similar perturbation analysis. But a homogeneous density profile is very restrictive and does not even fit with an isothermal self-gravitating cylinder (Ostriker 1964a). 
The calculation should ideally be applicable to any type of density profile and equation of state. 
Hansen et al. (1976) examined the case of an infinite, isothermal, and uniformly rotating cylinder of finite radius, whereas more complex situations were investigated from a numerical perspective (\cite{mitalas, bastien, arcoragi, bastien91, nakamura93, matsumoto, tomisaka95, nakamura, tomisaka96}). 

Nagasawa (1987) numerically obtained a dispersion relation in the case of an isothermal gas cylinder with an axial magnetic field, and found that such a cylinder was unstable to axisymmetric perturbations of wavelength higher than a particular one. This behavior is similar to the spherical Jeans case. 
Studying the stability of self-similar solutions for an infinitively long isothermal filament, Inutsuka \& Miyama (1992) found that filaments of line mass close to the critical value required for equilibrium were unstable to axisymmetric perturbations. 
When the line mass greatly exceeds the critical value, fragmentation does not occur and the entire filament globally collapses towards its axis.
Fischera \& Martin (2012) further characterized the gravitational state of isothermal gas cylinders with a finite radial boundary and a non-zero  external pressure in terms of the ratio of their line mass to the maximum possible value for a cylinder in vacuum. This maximum corresponds to the critical value of Inutsuka \& Miyama (1992) when the effective temperature is equal to the thermal temperature. The comparison of their results with observations of interstellar filaments indicated good agreement and suggested a temperature of 10 K.

Interest in the growth of instabilities within filaments has been recently revived, motivated by observations and simulations, and with the need for a stability analysis compatible with more realistic densities and equations of state. Notably, Quillen \& Comparetta (2010) roughly
estimated a dispersion relation inspired by Fridman \& Polyachenko (1984) in the case of the tidal tail of a galaxy, followed by Schneider \& Moore (2011), and Breysse et al. (2013)  carried out a detailed perturbation analysis for polytropic but non-rotating filaments.

\subsection{Current motivation and hypotheses}

Recent observations have shown that interstellar filaments are not isothermal, finding dust temperatures that tend to increase with radius (Stepnik et al. 2003, Palmeirim et al. 2013), and theoretical calculations inspired by Ostriker (1964) show that such filaments are more stable than their isothermal counterparts (\cite{recchi}). 
This motivates us not to limit ourselves to the isothermal case, but to assume a more general polytropic equation of state, which can result in a temperature varying with radius. For a given adiabatic index, not all density profiles would reproduce the observed trend for the temperature, 
but  so as to keep the calculation as general as possible, we do not restrict ourselves to any specific density distribution. 
The pressure support could be due to thermal as well as turbulent pressure. In fact, in the interstellar medium the dominant term is the turbulent support (\cite{mckee, hennebellefalg}).

Our idealized picture consists of an infinite self-gravitating and rotating cylinder with pressure and density related by a polytropic equation of state. 
We neglect the role of magnetic fields for simplicity, although their influence may contribute to the stabilization of the filaments (\cite{hennebelle}). 
We treat matter as an inviscid fluid and use hydrodynamics to obtain the linearized equations that govern the local perturbations. 
Cylindrical symmetry involves no dependance on the axial and azimuthal coordinates $z$ and $\phi$ for the unperturbed system, and we only consider axisymmetric perturbations. We further assume that there is no radial movement in the undisturbed system and that all particles initially share the same axial velocity. 
As filamentary geometry provides a favorable situation for small perturbations to grow before global collapse overwhelms them (Pon et al. 2011), we assume an unperturbed system at equilibrium and neglect the global collapse of the filamentary cloud.

\section{Obtention of the dispersion relation}

\subsection{The unperturbed system}

The unperturbed system is assumed to be infinite and cylindrically symmetric.
The undisturbed density is thus written as $\rho_0 (R)$, pressure as $p_0 (R)$, and the resulting gravitational field as $\Phi_0 (R)$,  in cylindrical coordinates. 
We assume that there is no radial velocity in the undisturbed system and that all fluid particles share the same initial axial velocity. The velocity field can then be written in the reference frame of the cylinder as $\vec{\mathrm{v}_0} (R,\phi) = R \Omega_0 (R)~\vec{e_\phi}$, where $\Omega_0 (R)$ is the undisturbed angular velocity and $\vec{e_\phi}$ the azimuthal unit vector. This expression guarantees mass conservation, and all calculations will hereafter be carried in the reference frame of the unperturbed system, without loss of generality. 

For the unperturbed system, the radial equation of motion and the Poisson equation can be written as
\begin{equation}
\label{eq:1}
 - R \Omega_0^2 = - \frac{\partial h_0}{\partial R} - \frac{\partial \Phi_0}{\partial R}
\end{equation}
\begin{equation}
\label{eq:2}
 \nabla^2 \Phi_0 = 4 \pi G \rho_0
\end{equation}
where the enthalpy $h_0 (R)$ is defined by $dh_0 = \rho_0^{-1} dp_0$. The polytropic equation of state relates the pressure $p_0$ to the density $\rho_0$  by a simple power law. 

\subsection{Linear perturbations}

We study the growth of instabilities when the physical quantities deviate from their equilibrium values. 
The dynamics of the perturbed system is  determined by the following set of linearized first-order equations, where the infinitesimal disturbances are denoted by an index 1: 
\begin{equation}
	 \displaystyle \frac{\partial \vec{\mathrm{v}_1}}{\partial t} + \left(\vec{\mathrm{v}_0}.\vec{\nabla}\right) \vec{\mathrm{v}_1} +  \left(\vec{\mathrm{v}_1}.\vec{{\nabla}}\right) \vec{\mathrm{v}_0}  = -\vec{{\nabla}} h_1 - \vec{{\nabla}} \Phi_1
\end{equation}
\begin{equation}
	 \displaystyle \frac{\partial \rho_1}{\partial t} + \mathrm{div} \left(\rho_1 \vec{\mathrm{v}_0}\right)+ \mathrm{div} \left(\rho_0 \vec{\mathrm{v}_1}\right)  = 0
\end{equation}
\begin{equation}
	 \displaystyle \nabla^2 \Phi_1 = 4 \pi G \rho_1.
\end{equation}
These equations correspond respectively to the equation of motion, the continuity equation, and the Poisson equation. The linearized equation of state  yields for its part  (see Appendix \ref{appendix:enthalpy})
\begin{equation}
	 \displaystyle h_1 = c_0^2~ \frac{\rho_1}{\rho_0}
\end{equation}
where $c_0 (R)$ is the speed of sound, defined by $c_0 ^2 = \partial p_0/\partial \rho_0$. For an isothermal cylinder, $c_0$ would be a constant related to the gas temperature, but the following calculations apply to any polytropic equation of state. Assuming axisymmetric perturbations of the generic form $ e^{-i\omega t} e^{ik_R R} e^{i k_z z}$ (normal modes), we obtain
\begin{equation}
	 \displaystyle \omega \mathrm{v}_{1R}- 2i \Omega_0 \mathrm{v}_{1\phi} = - i \frac{\partial h_1}{\partial R} + k_R \Phi_1
\end{equation}
\begin{equation}
	 \displaystyle \omega \mathrm{v}_{1\phi} - 2 i B \mathrm{v}_{1R} = 0
\end{equation}
\begin{equation}
	 \displaystyle \omega \mathrm{v}_{1z} = k_z c_0^2 \frac{\rho_1}{\rho_0} + k_z \Phi_1
\end{equation}
\begin{equation}
	 \displaystyle \omega \rho_1 + i \frac{1}{R}\frac{\partial}{\partial R} \left(R\rho_0 \mathrm{v}_{1R}\right)-k_z \rho_0 \mathrm{v}_{1z} = 0
\end{equation}
\begin{equation}
	  \displaystyle \frac{1}{R}\frac{\partial}{\partial R}\left(R \frac{\partial\Phi_1}{\partial R}\right)-k_z^2 \Phi_1 = 4\pi G \rho_1
\end{equation}
where the equation of motion has been projected on the different cylindrical axes, and $ B(R) = - \frac{1}{2} \left(\Omega_0 (R) +\frac{\partial}{\partial R}\left(R \Omega_0(R)\right)\right)$ is one of Oort constants (\cite{jog}). 

\subsection{Local perturbations}
\label{disprel}

We assume local perturbations: the typical scale of the perturbation is small compared to that of the unperturbed quantities, i.e., $k_R R_0 >> 1$, where $R_0$ is the typical radius for the unperturbed distribution within which the perturbation occurs. This assumption was made in a seminal article by Toomre (1964), and is analogous to the Wentzel-Kramers-Brillouin approximation (WKB) used in quantum physics. Perturbed quantities $X_1$ oscillate rapidly with the radius $R$ whereas unperturbed quantities $X_0$ change much more smoothly along the radial direction, so that 
\[
\displaystyle \left|X_1 \frac{\partial X_0}{\partial R}\right| << \left|X_0 \frac{\partial X_1}{\partial R}\right|.
\]
This approximation yields up to the first order
\begin{equation}
\frac{\partial h_1}{\partial R} = \frac{\partial}{\partial R}\left(\frac{c_0^2}{\rho_0} \rho_1\right) = \frac{c_0^2}{\rho_0}\frac{\partial \rho_1}{\partial R} + \rho_1\frac{\partial}{\partial R}\left(\frac{c_0^2} {\rho_0}\right) \simeq \frac{c_0^2}{\rho_0}\frac{\partial \rho_1}{\partial R}
\end{equation}
\begin{equation}
\frac{1}{R}\frac{\partial}{\partial R} \left(R \rho_0 \mathrm{v}_{1R}\right) = \rho_0 \frac{\partial \mathrm{v}_{1R}}{\partial R} + \mathrm{v}_{1R} \frac{1}{R}\frac{\partial}{\partial R}\left(R\rho_0\right) \simeq \rho_0 \frac{\partial \mathrm{v}_{1R}}{\partial R}
\end{equation}
\begin{equation}
\frac{1}{R}\frac{\partial}{\partial R}\left(R \frac{\partial\Phi_1}{\partial R}\right) = \frac{\partial^2 \Phi_1}{\partial R^2} + \frac{1}{R}\frac{\partial\Phi_1}{\partial R} \simeq \frac{\partial^2 \Phi_1}{\partial R^2}
\end{equation}
and the linearized equations become
\begin{equation}
	   \omega \mathrm{v}_{1R}- 2i \Omega_0 \mathrm{v}_{1\phi} = k_R \frac{c_0^2}{\rho_0}  \rho_1 + k_R \Phi_1
\end{equation}
\begin{equation}
	   \omega \mathrm{v}_{1\phi} - 2 i B \mathrm{v}_{1R} = 0
\end{equation}
\begin{equation}
	   \omega \mathrm{v}_{1z} = k_z \frac{c_0^2}{\rho_0} \rho_1 + k_z \Phi_1
\end{equation}
\begin{equation}
	   \omega \rho_1 - \rho_0 k_R \mathrm{v}_{1R}-\rho_0 k_z \mathrm{v}_{1z} = 0
\end{equation}
\begin{equation}
	    -k_R^2 \Phi_1-k_z^2 \Phi_1 = 4\pi G \rho_1.
\end{equation}
Combining these equations, we obtain the local dispersion relation, 
\begin{equation} 
\label{eq:disp}
\omega^4 + \omega^2 \left( 4\pi G \rho_0 - c_0^2 k^2-\kappa^2\right) + \kappa^2 k_z^2 \left(c_0^2-\frac{4\pi G \rho_0}{k^2} \right) = 0
\end{equation}
where $k = \sqrt{\smash{k_R}^2 +\smash{k_z}^2}$ corresponds to the total wavenumber and $\kappa$ is the epicyclic frequency, defined by $\kappa^2 = - 4 \Omega_0 B$. Note that this latter quantity, as well as $\rho_0$ and $c_0$, does depend on the position $R$ at which the perturbation is considered.

When there is no rotation, $\kappa^2 = 0$ and the asymmetry between the different directions disappears at short scales. Eq. (\ref{eq:disp}) then reduces to the standard dispersion relation for collapsing spherical systems (\cite{binney}):
\begin{equation}
 \omega^2 = c_0^2 k^2 - 4\pi G \rho_0.
\end{equation}
We thus retrieve the Jeans criterion, with perturbations larger than $\lambda_J = (\pi c_0^2 / G\rho_0)^{1/2}$ being unstable, where $\rho_0$ and $c_0$ still depend on the position $R$.

\section{Some properties of the dispersion relation}
\label{properties}

Here we derive some properties of the dispersion relation, without assuming a specific density profile. 

\subsection{A condition for stability}
\label{cond}

\noindent Introducing the characteristic angular frequency $\omega_0 = \sqrt{4\pi G \rho_c}$, where $\rho_c = \rho_0 (0)$ is the central density of the filament, the dispersion relation obtained in section \ref{disprel} (Eq. (\ref{eq:disp})) can be written in terms of the dimensionless variable $ x = \omega/\omega_0$ as a fourth order polynomial equation: 
\begin{equation}
\label{eq:dispx}
 x^4 + \alpha x^2 + \beta = 0
\end{equation}
where $\displaystyle \alpha = \frac{\rho_0}{\rho_c} - \frac{c_0^2 k^2}{\omega_0^2} - \frac{\kappa^2}{\omega_0^2}$ and 
$ \displaystyle\beta =  \frac{\kappa^2 k_z^2}{\omega_0^2} \left(\frac{c_0^2}{\omega_0^2} - \frac{1}{k^2} \frac{\rho_0}{\rho_c}\right)$. 
This equation can also be considered as a second order polynomial expression in $x^2$, whose roots are $x_+^2 = \frac{-\alpha + \sqrt{\Delta}}{2}$ and $x_-^2=\frac{-\alpha - \sqrt{\Delta}}{2}$, where $\Delta  = \alpha^2 - 4\beta$ is the discriminant of the equation. 
It can be shown (Appendix \ref{appendix:discriminant1}) that $\Delta$ is always positive, implying that $x^2$ is always real. Moreover, $x_+^2$ is always positive as well (Appendix \ref{appendix:discriminant2}), so at least two of the four roots of Eq. (\ref{eq:dispx}) are real. Consequently, the system is  stable to axisymmetric perturbations when $x_-^2 ~\geqslant~0$ and unstable when $x_-^2 < 0$, as growing modes require a non-zero imaginary part. 
This condition reduces to the standard Jeans criterion (Appendix \ref{appendix:discriminant}); i.e., all roots are real and the system is stable when 
\begin{equation}
\label{eq:jeans}
 k^2 \geqslant 4\pi G \rho_0/c_0^2.
\end{equation}
Introducing the critical wavenumber $k_{\mathrm{crit}} = \sqrt{4\pi G \rho_0}/c_0$, the filament is expected to be stable for wavenumbers $k \geqslant k_{\mathrm{crit}}$ and unstable below.
This stability condition is symmetrical in $k_R$ and $k_z$, does not depend on rotation, and depends on the radius $R$ at which the perturbation occurs as $k_{\mathrm{crit}}$ is a function of $\rho_0$ and $c_0$. 
Symmetry arises because of the local assumption, which prevents the space-limited perturbation feeling the larger scale behavior of the system. 
The result is consistent with observations of structures within an interstellar filament by Kainulainen et al. (2013), which show that the fragmentation more closely ressembles Jeans fragmentation at small scales than at higher scales. At small scales, fragmentation is insensitive to the large-scale geometry and only depends on local properties. 

As this equation was established for a polytropic equation of state where $c_0$ could vary with  radius, it can also describe filaments with a non-uniform temperature. Palmeirim et al. (2013) are notably able to reproduce the increasing temperature profile of a filament in the Taurus molecular cloud with a polytropic equation of state and a Plummer-like density profile, so that the temperature is proportional to $\rho_0(R)^{\gamma -1}$ with an adiabatic index $\gamma < 1$. 
For such filaments with temperature and sound speed increasing with radius, the critical wavenumber $k_{\mathrm{crit}}$ falls more rapidly with radius than in the isothermal case. Consequently, such filaments are expected to be more stable than their isothermal counterparts, which was also inferred by Recchi et al. (2013) by comparing the equilibrium structure of isothermal and non-isothermal filaments.

In the isothermal case, Nagasawa (1987) numerically computes the dispersion relation for axisymmetric perturbations within an infinite non-rotating gas cylinder, and obtains a critical wavenumber that is 0.561 the critical wavenumber derived from Eq. (\ref{eq:jeans}) at the center of a similar filament. 
The discrepancy is partly due to the simplification involved in the local assumption, and thus gives an idea of its limitations.

\subsection{The fastest growing mode}
\label{fastest}

The only solution of the dispersion relation (Eq. (\ref{eq:disp})) that could be negative and give rise to unstable modes is $ x_-^2$. 
For a given radius $R$, the distribution of $x_-^2$ in the phase space ($k_R$, $k_z$) is a useful tool to study the properties of the instabilities that could form. As long as $k<k_{\mathrm{crit}}$, different unstable modes can grow and coexist. But the fastest growing mode, which minimizes the value of $x_-^2$ for a given $k_R$, will dominate the growth of instabilities and influence the shape of the resulting structures. At any given radius $R$, this fastest growing mode can be described by a polynomial equation in $k^2$ and $k_R^2$ (Appendix~\ref{appendix:fastest1}):
\begin{equation}
\left(\frac{\rho_0}{\rho_c}\right)^2 \frac{k_R^2}{k^2} =
\frac{c_0^2}{\kappa^2} k^2 \left[ \left(\frac{c_0^2}{\omega_0^2} k^2 - \frac{\rho_0}{\rho_c}\right)^2+ \frac{\rho_0}{\rho_c}\frac{\kappa^2}{\omega_0^2}\right].
\end{equation}
For a given $k_R$, this equation can be rewritten as a polynomial expression in $k/k_0$, where $k_0= \omega_0/c_0$ is a characteristic wavenumber, depending in principle on the position $R$ through $c_0$: 
\begin{equation}
\label{eq:curve}
\left(\frac{k}{k_0}\right)^8{-}2\frac{\rho_0}{\rho_c}\left(\frac{k}{k_0}\right)^6{+}\frac{\rho_0}{\rho_c}\left( \frac{\rho_0}{\rho_c}{+}\frac{\kappa^2}{\omega_0^2}\right)\left(\frac{k}{k_0}\right)^4{-}\frac{\kappa^2}{\omega_0^2}\left(\frac{\rho_0}{\rho_c}\right)^2 \left(\frac{k_R}{k_0}\right)^2{=}0.~
\end{equation}
For an isothermal filament, $k_0$ corresponds to the critical wavenumber taken at the center of the filament.
When the curve associated with Eq. (\ref{eq:curve}) in the ($k_R$, $k_z$) plane is above the line $k_z = k_R$, 
the structure resulting from the perturbation is expected to be oblate, whereas when $k_z<k_R$, the structure is expected to be prolate. 
This curve meets the line $k_z = k_R$ when $k_R$ satisfies (Appendix~\ref{appendix:fastest2}):
\begin{equation}
\label{eq:kr}
 \left(\frac{k_R}{k_0}\right)^6-\frac{\rho_0}{\rho_c} \left(\frac{k_R}{k_0}\right)^4 + \frac{1}{4} \frac{\rho_0}{\rho_c} \left(\frac{\rho_0}{\rho_c}+\frac{\kappa^2}{\omega_0^2}\right) \left(\frac{k_R}{k_0}\right)^2 - \frac{1}{16} \left(\frac{\rho_0}{\rho_c}\right)^2 \frac{\kappa^2}{ \omega_0^2} =0.~~
 \end{equation}
This polynomial equation can be solved numerically, and should give a threshold value for $k_R$ separating prolate and oblate collapsed structures. 

\subsection{A representation of the dispersion relation for an interstellar filament (TMC-1, or ``The Bull's Tail'')}
\label{section:tmc}

\begin{figure*}
\centering
   \includegraphics[width=17cm]{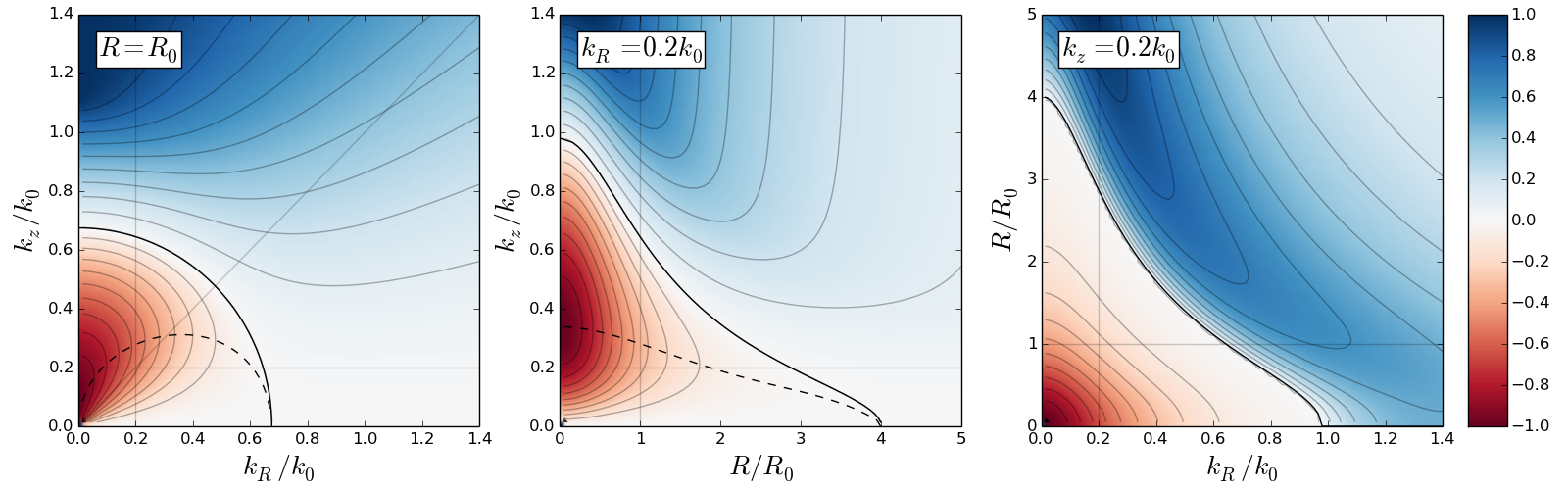}
     \caption{Projection of the distribution of the frequency $x_-^2$ in the planes $R = R_0$, $k_R = 0.2 k_0$, and $k_z = 0.2 k_0$ for an idealized filament inspired by TMC-1, with $R_0 = 0.043$ pc and $k_0 = 74$ pc$^{-1}$. The color scale is normalized to its maximum for positive values of $x_-^2$ and to its minimum for negative values. The contours are equally spaced. Negative values of $x_-^2$ correspond to an unstable filament, whereas the filament should be stable in the region where $x_-^2$ is positive. The solid black curve separates the stable and unstable regimes, and corresponds to $k = k_{\mathrm{crit}}$. 
This threshold corresponds to a length scale $\lambda_{\mathrm{crit}} = 2\pi / k_{\mathrm{crit}} =  0.13$ pc at $R=R_0$. The dashed line corresponds to the minimum value of $x_-^2$, i.e., to the most unstable mode, and is obtained from Eq. (\ref{eq:curve}) for the first panel. It intersects the line $k_z = k_R$ at $k_{R,\mathrm{threshold}} = 0.31 k_0$, which corresponds to $\lambda_{\mathrm{threshold}} = 0.28$~pc.
}
     \label{tmc}
\end{figure*}

As an illustrative numerical application, we use a filament from the Taurus molecular cloud, TMC-1, also known as ``The Bull's Tail'' (\cite{nutter}). 
Recent \textit{Herschel} Gould Belt observations were able to reveal the structure of interstellar filaments with unprecedented detail (\cite{andre}, \cite{arzoumanian}, \cite{malinen}, \cite{palmeirim}, \cite{kirk}), and
the Taurus molecular cloud is one of the closest and most studied star-forming regions (e.g., \cite{narayanan, goldsmith}). 
Malinen et al. (2012) fit the density profile of this filament with a Plummer-like profile of the form 
\begin{equation}
\displaystyle \rho_0 (R) =  \frac{\rho_c}{\left(1+\left(\frac{R}{R_0}\right)^2\right)^p}
\end{equation}
and obtain p = 1.1, $R_0 = 0.043$ pc, and $\rho_c = 7.0 \times 10^{4} ~\mathrm{cm}^{-3}$. 
We consider an idealized filament whose density profile corresponds to the analytical Plummer-like profile found by Malinen et al. (2012), and numerically solve the dispersion relation (Eq. (\ref{eq:disp})). 
We use a uniform thermal velocity $c_0 = 200$ m.s$^{-1}$ for simplicity, which corresponds to a temperature of about 10 K. 
We obtain $x_-^2$ as a function of the wavenumbers $k_R$ and $k_z$ and of the radius $R$ at which the perturbation is assumed to arise. $ x_-^2$ is indeed the only solution of the dispersion relation that could be negative and give rise to unstable modes. 
Figure \ref{tmc} shows the resulting distribution of $x_-^2$ projected in the planes $R = R_0$, $k_R = 0.2 k_0$, and $k_z = 0.2 k_0$, where $k_0 = \omega_0/c_0$ is the characteristic wavenumber, here equal to $74$ pc$^{-1}$. As expected from Eq. (\ref{eq:jeans}), while the boundary between the stable and unstable regimes is symmetrical in the plane ($k_R$, $k_z$), rotation and geometry introduce strong asymmetries in the unstable regime. These asymmetries could affect the shape of the prestellar cores that form within the filament. 
Notably, as the dashed line corresponds to the fastest growing instability, in the unstable regime we expect perturbations with a small $k_R$ to be less elongated in the $z$ direction 
than perturbations with a larger $k_R$. 
As shown by the figure and Eq. (\ref{eq:kr}), this curve intersects the line $k_z = k_R$ around $k_{R,\mathrm{threshold}}/k_0 = 0.31$. We thus expect to observe oblate structures of radius greater than $\lambda_{\mathrm{threshold}} = 2\pi / k_{R, \mathrm{threshold}} = 0.28$ pc, and prolate structures below. 

Comparing this result with observations of protostellar regions or molecular clouds is not straightforward, as most studies are more interested in the statistical distribution of shapes than in the correlation between shape and size, and as projected quantities are not fully reliable tracers of three dimensional shapes. 
Most observations in the Taurus molecular cloud or in other molecular clouds favor prolate structures within interstellar filaments and tend to show that cores are elongated along the direction of the filaments  (e.g., \cite{myers91, nozawa, tatematsu, onishi, tachihara, curry, hartmann}), which is expected in a picture where cores are formed by gravitational fragmentation of their host filament (e.g., \cite{schneider}). 
Jones et al. (2001) and Jones~\&~Basu~(2002) also use catalogs of observations for molecular clouds, cloud cores, Bok globules, and condensations to characterize their shape and show that neither pure oblate nor pure prolate cores can account for the observed distribution, and that cores have intrinsically triaxial shapes, but tend to be more prolate on scales $\gtrsim$~1 parsec and closer to oblate between 0.01 and 0.1 parsec. 
Our result yields the opposite trend, but the comparison between the two studies may not be straightforward, and using projected axis ratios may not be the best method to distinguish between oblate and prolate structures, as it introduces systematic biases (\cite{curry}).

As our calculations are based on the local assumption $k_R R_0 >> 1$, the dispersion relation does not rigorously apply when the wavelength $2\pi/k_R$ is large compared to the typical size of the density distribution, and it is thus difficult to assess the relative importance of oblate and prolate cores from the dispersion relation (Eq. (\ref{eq:disp})). 
Moreover, observation of filaments with clumps of star formation whose size is comparable to the thickness of the filament could imply that the whole filament collapses along its axis into several clumps, which would not be a local collapse.

The discrepancies between the predicted and observed shapes could not only be the consequence of non-local perturbations, but could also be explained if observed structures have grown in the non-linear regime. A fastest growing mode that is an oblate perturbation in the linear regime could further collapse to a more prolate structure within the embedding filament in the non-linear regime, as suggested by the numerical calculation by Inutsuka \& Miyama (1997) for isothermal filaments of line mass close to the critical equilibrium value.
Rotation also logically tends to favor oblate perturbations, thus underestimating the velocity scale appearing in the pressure term would necessarily produce more oblate structures than it should. 
Further studies would be necessary to evaluate the match between observations and the dispersion relation we obtain, which is beyond the illustrative scope of this numerical application.

\subsection{An order-of-magnitude estimate for cosmic filaments}

Maps of the galaxy distribution in the Universe reveal large coherent structures such as filaments and walls (e.g., \cite{geller, gott}). Simulations explain these alignments and predict the existence of a dark matter and gaseous cosmic web that would connect galaxies one to another (e.g., \cite{bond, springel, keres, dekel2009, fumagalli}). 
Direct evidence of the dark matter and gas filaments is still lacking, and we have to rely on simulations to describe them. Harford \& Hamilton (2011) study a cosmological simulation at redshift $z= 5$ and show that the inner core of intergalactic filaments is predominantly made of gas and that a significant fraction of them can be described as isothermal gas cylinders. The density profile of an infinite, self-gravitating, isothermal gas cylinder in hydrodynamical equilibrium is given by 
\begin{equation}
\displaystyle \rho_0 (R) =  \frac{\rho_c}{\left(1+\frac{1}{8}\left(\frac{R}{R_s}\right)^2\right)^2}
\end{equation}
with $R_s = c_s/\sqrt{4\pi G \rho_c}$ (\cite{ostriker}). An effective sound speed $c_s$ can thus be computed from the isothermal profile, and this quantity is shown to correspond to the actual sound speed determined from the temperature for a signifiant fraction of the filaments. For filaments whose parent shell overdensity is above 10 and whose gas fraction is higher than 50\%, the fraction of isothermal filaments with an effective sound speed equal within a factor 2 to the actual sound speed indeed reaches 50\%. The central gas density of these filaments is about 500 times the mean cosmic density at the simulation redshift, i.e., about 0.1 cm$^{-3}$, and the temperature of 12~000-14~000~K corresponds to a sound speed of about 7 km s$^{-1}$. The stability threshold given by Eq. (\ref{eq:jeans}) applied at the center of such a filament leads to a characteristic scale $\lambda_{\mathrm{crit}}\simeq 3$ kpc for the structures that would form within it, which corresponds to the size of the filaments. 
This is consistent with a model in which cosmic filaments are shaped by gravity and are not merely intersections of sheets. In retrospect, this justifies  modelling them as self-gravitating cylinders. 

\section{Conclusion and discussion}

Assuming perturbations smaller than the typical size of the density distribution, we derived a dispersion relation for axisymmetric perturbations in an infinite, self-gravitating, and rotating filament (Eq. (\ref{eq:disp})). The gas is assumed to be polytropic - which includes the narrower isothermal case - and the relation is valid for any type of density profile. 
This dispersion relation yields a symmetrical threshold in the axial and radial wavenumbers between the stable and unstable regions, but the fastest growing mode breaks this symmetry and should influence the shape of the resulting perturbations. 
We used an interstellar filament observed in the Taurus Molecular Cloud, TMC-1, as fiducial numerical application, and represented the properties of the dispersion relation in the ($k_R$, $k_z$) phase space as well as in the ($R$, $k_R$) and ($R$, $k_z$) planes. Perturbations of radius greater than a characteristic length are expected to lead to oblate structures, and to prolate structures below. 
Simulations of cosmic filaments are compatible with a gravitational origin and validate our assumption to model them as self-gravitating cylinders.

Gravitational collapse should enhance rotation in interstellar and cosmic filaments due to angular momentum conservation. But although signs of rotation such as transverse velocity gradients are observed for interstellar filaments, and notably for TMC-1 (\cite{olano}), there generally does not seem to be a global coherent rotation of such filaments, as assumed in our calculations (e.g., \cite{loren, tatematsu, falgarone01}). Hence, our calculations may overestimate the effects of rotation, and thus artificially favor oblate structures. 
Nevertheless, the consequences of a turbulent velocity dispersion varying with radius should be studied more carefully, as the resulting stability gain against gravitational collapse could replace that due to rotation, with similar scale dependence according to density distribution.
The velocity dispersion $\sigma$ is indeed expected to increase with the size $L$ of the filament, as given by Larson's relation $\sigma \propto L^{0.5}$ (\cite{larson, solomon}), and would thus supply an additional support against gravitational collapse for larger radii.

We neglected the role of magnetic fields in our calculations, although they are ubiquitous in the interstellar medium and could affect filament stability. Even a small magnetic field can generally play a significant role in gas dynamics, and moderate fields can have a strong impact on the fragmentation of gas clouds and on the formation of prestellar cores (\cite{tilley, li}). The magnetic forces are indeed intrinsically anisotropic, which promotes fragmentation. 
By studying numerically the dispersion relation for magnetized and non-magnetized isothermal filaments, Nagasawa (1987) has shown that a uniform axial magnetic field does not change the critical wavenumber but efficiently stabilizes the filament by reducing the growth rate of the unstable modes. 
Observationally, dense self-gravitating filaments tend to appear perpendicular to the direction of the local magnetic field, whereas their lower-density unbound striations tend to be parallel to it (\cite{andre14}). 
A generalization of our calculations including magnetic fields in different configurations would help us understand the formation of prestellar cores within interstellar filaments. 

Our model also implicitly assumes an isolated filament, surrounded by voids, whereas interstellar and cosmic filaments are typically embedded in intricate networks. Indeed, interstellar filaments tend to branch out from dense star-forming hubs (\cite{myers2009}) or to group in smaller-scale bundles of similar properties and common physical origin (\cite{hacar}). These bundles may result from the fragmentation of the initial cloud into different sub-regions, that further condense into velocity-coherent filaments. Gas accretion onto the filaments is also expected to play a major role, as it not only brings mass but also drives internal turbulence and affects the stability of the filament (\cite{klessen, heitsch}).  
The interaction between filaments and with their environment, including tidal fields as shown in the spherical Jeans case (\cite{jog2013}), is likely to influence their fragmentation, and should be further investigated. 

While the study of the environmental effects on the formation of prestellar cores requires a more detailed study, some limited generalizations of our calculations could be achieved more easily. 
We notably limited ourselves to axisymmetric perturbations, while the case of non-axisymmetric perturbations could also be of interest. Such perturbations are notably envisaged numerically for an isothermal cylinder by Nagasawa (1987), which could provide a useful comparison. 
%
We assumed the same initial axial velocity for all particles, which enabled us to remove the velocity dependence in the reference frame of the unperturbed system. Although there may be an average axial flow towards one edge of the filament, the initial velocity distribution of the particles is likely to be more complex than assumed here. For example, the case of a Maxwellian distribution could be studied in more detail, as well as the case of accelerated particles. Peretto et al. (2013) and Zernickel et al. (2013) indeed observe coherent velocity gradients along interstellar filaments and interpret them as a large-scale longitudinal collapse. Filaments feel the gravitational acceleration from the structures onto which they are accreted, thus leading to accelerated particles. 
The longitudinal expansion of tidal tails has notably been shown to have a strong stabilizing effect (\cite{schneider11}).
%
We also considered the analytical case of a polytropic equation of state. More complex equations of state could be envisaged, notably a barotropic one. This last type of equation of state has the advantage of being more general than a polytropic one, while remaining easily described mathematically. 
We plan to follow up some of these ideas in future papers, as well as address the influence of magnetic fields on our results, and carry out a more detailed comparison of the local dispersion relation found here with observations and simulations. 

\begin{acknowledgements}
This publication benefited from the European Research Council Advanced Grant Program number 267399 - Momentum, 
and J.F. acknowledges support from the Indo French Centre for the Promotion of Advanced Research (IFCPAR/CEFIPRA) through a Raman-Charpak fellowship.
The authors wish to thank the anonymous referee, whose comments have led to significant improvements in this paper, and Martin Stringer for the proofreading. 
\end{acknowledgements}


\newpage

\begin{appendix} 
\section{Expressions of the enthalpy}
\label{appendix:enthalpy}

We assume a polytropic equation of state $p_0 = C \rho_0^\gamma$, where $C$ is a positive constant and $\gamma$ the adiabatic index. The speed of sound is defined by $c_0 ^2 = \partial p_0/\partial \rho_0 = \gamma C \rho_0 ^{\gamma-1} = \gamma p_0/\rho_0$, so that when $\gamma \neq 1$, the enthalpy can be written as:
\begin{equation}
h_0 = \frac{\gamma}{\gamma -1} C \rho_0^{\gamma-1} = \frac{c_0^2}{\gamma -1}.
\end{equation}
In the case of an isothermal equation of state ($\gamma = 1$), the speed of sound is constant and entirely fixed by temperature.  
The isothermal equation of state can be written $p_0 = c_0^2~\rho_0$ and the enthalpy yields:
\begin{equation}
h_0 = c_0^2 ~\mathrm{ln} \rho_0.
\end{equation}
In both cases, the first order perturbation of the enthalpy can be written as:
\begin{equation}
h_1 = c_0^2~ \frac{\rho_1}{\rho_0}.
\end{equation}
Indeed, the perturbed fluid remains polytropic, so up to the first order in $\rho_1/\rho_0$, when $\gamma \neq  1$, 
\[
h_0 + h_1 = \frac{\gamma C}{\gamma-1} ~ \left(\rho_0+\rho_1\right)^{\gamma-1} = h_0 + \gamma C~ \rho_0^{\gamma-2}\rho_1 = h_0 + c_0^2  \frac{\rho_1}{\rho_0}
\]
and when $\gamma = 1$, 
\[
h_0 + h_1 =  c_0^2~ \mathrm{ln} \left(\rho_0+\rho_1\right) = h_0 + c_0^2  \frac{\rho_1}{\rho_0}.
\]

\end{appendix}

\begin{appendix} 
\section{Discriminant of the dispersion relation}
\label{appendix:discriminant}

\subsection{The discriminant is always positive}
\label{appendix:discriminant1}

The dimensionless dispersion relation introduced in section \ref{properties} (Eq. (\ref{eq:dispx})) involves two dimensionless quantities, $\alpha$ and $\beta$. Considering that 
\begin{equation}
\label{eq:alpha}
\alpha = -\left(\frac{\omega_0^2 k^2}{\kappa^2 k_z^2} \beta + \frac{\kappa^2}{\omega_0^2}\right)
\end{equation}
the discriminant can be written as a second order polynomial expression in $\beta$: 
\begin{equation}
\Delta  = \displaystyle \alpha^2 - 4\beta = \displaystyle \frac{\omega_0^4 k^4}{\kappa^4 k_z^4} \beta^2 +2 \frac{k_R^2-k_z^2}{k_z^2} \beta + \frac{\kappa^4}{\omega_0^4}.
\end{equation}
In turn, the discriminant $\Delta^\prime$ of this latter polynomial expression is always negative: 
\begin{equation}
\Delta^\prime =  \displaystyle \left(\frac{k_R^2-k_z^2}{k_z^2}\right)^2 - \frac{k^4}{ k_z^4} 
= - 4 \left(\frac{k_R}{k_z}\right)^2.
\end{equation}
Consequently, as $\beta$ is a real quantity, the discriminant $\Delta$ is always positive and $ x_\pm^2 = \frac{-\alpha \pm \sqrt{\Delta}}{2}$  always real. 

\subsection{Two roots are real}
\label{appendix:discriminant2}

There are four solutions for $x$, which can be either real or with a non-zero imaginary part, depending on the sign of $-\alpha \pm \sqrt{\Delta}$. But out of these four solutions, two are always real. Indeed, $-\alpha+\sqrt{\Delta}$ is always positive:
\begin{itemize}
\item If $\alpha \leqslant 0$, $-\alpha+\sqrt{\Delta} \geqslant 0$.\\
\item If $\alpha > 0$, $\beta$ has to be negative because of Eq. (\ref{eq:alpha}), so \mbox{$\Delta=\alpha^2-4\beta > \alpha^2$} and $-\alpha+\sqrt{\Delta} > 0$.
\end{itemize}

\subsection{A condition for stability}
\label{appendix:discriminant3}

All roots are real and the system is stable when and only when $-\alpha-\sqrt{\Delta}$ is also positive, i.e., when $\alpha \leqslant 0$ and $\left|\alpha\right| \geqslant \sqrt{\Delta}$. These two conditions can be expressed as conditions on the total wavenumber $k$: 
\begin{equation}
\begin{array}{lcl}
\displaystyle  \alpha \leqslant 0 & \displaystyle \Leftrightarrow & \displaystyle \frac{\rho_0}{\rho_c} - \frac{c_0^2 k^2}{\omega_0^2} - \frac{\kappa^2}{\omega_0^2} \leqslant 0\\
& \displaystyle \Leftrightarrow & \displaystyle k^2 \geqslant \frac{4\pi G \rho_0-\kappa^2}{c_0^2}
\end{array}
\end{equation}
\begin{equation}
\begin{array}{lcl}
\displaystyle \left|\alpha\right| \geqslant \sqrt{\Delta} & \displaystyle \Leftrightarrow & \alpha^2 \geqslant \alpha^2-4\beta\\
& \displaystyle \Leftrightarrow & \displaystyle \beta \geqslant 0\\
& \displaystyle \Leftrightarrow & \displaystyle \frac{c_0^2}{\omega_0^2} - \frac{1}{k^2} \frac{\rho_0}{\rho_c} \geqslant 0\\
& \displaystyle \Leftrightarrow & \displaystyle k^2 \geqslant \frac{4\pi G \rho_0}{c_0^2}.
\end{array}
\end{equation}
The second condition encompasses the first one, and is analogous to the Jeans criterion. All roots are thus real and the system is stable for $\displaystyle k \geqslant k_{\mathrm{crit}}$, with $k_{\mathrm{crit}}^2 =  4\pi G \rho_0/c_0^2$. 

\end{appendix}

\begin{appendix} 
\section{About the fastest growing mode}
\label{appendix:fastest}

\subsection{A parametric expression for the fastest growing mode}\label{appendix:fastest1}

\noindent At a given radius $R$ and for a given $k_R$, the quantity $x_-^2 = \frac{-\alpha - \sqrt{\Delta}}{2}$ introduced in section \ref{cond} is minimal when 
\begin{equation}
\begin{array}{ll}
\displaystyle \frac{\partial x_-^2}{\partial k_z} = 0  & \displaystyle \Leftrightarrow~~~ \frac{\partial}{\partial k_z} \left(\alpha+\sqrt{\Delta}\right) = 0\\
& \displaystyle \Leftrightarrow~~~  \left(1+\frac{\alpha}{\sqrt{\Delta}}\right)\frac{\partial \alpha}{\partial k_z} - \frac{2}{\sqrt{\Delta}} \frac{\partial \beta}{\partial k_z} = 0 \\
& \displaystyle \Leftrightarrow~~~ k_z \left(1+\frac{\alpha}{\sqrt{\Delta}}+\frac{2 \kappa^2}{c_0^2 \sqrt{\Delta}} \left(\frac{c_0^2}{\omega_0^2}- \frac{\rho_0}{\rho_c}  \frac{k_R^2}{k^4}\right)\right) =0\\
& \Leftrightarrow~~~ \displaystyle k_z = 0 \mathrm{~or~} \sqrt{\Delta}  = -\alpha - 2 \frac{\kappa^2}{c_0^2} \left(\frac{c_0^2}{\omega_0^2}- \frac{\rho_0}{\rho_c}  \frac{k_R^2}{k^4}\right).
\end{array}
\end{equation}
When $k_z \neq 0$, the latter condition yields
\begin{equation}
   \left\{ 
	\begin{array}{l}
	\displaystyle \alpha + 2 \frac{\kappa^2}{c_0^2} \left(\frac{c_0^2}{\omega_0^2}- \frac{\rho_0}{\rho_c}  \frac{k_R^2}{k^4}\right) \leqslant 0\\
	\mathrm{and} \\
	\displaystyle \Delta = \alpha^2-4\beta = \left(\alpha + 2 \frac{\kappa^2}{c_0^2} \left(\frac{c_0^2}{\omega_0^2}- \frac{\rho_0}{\rho_c}  \frac{k_R^2}{k^4}\right)\right)^2.\\
	\end{array}
  \right.
\end{equation}
Using the expressions of $\alpha$ and $\beta$ and developing the intermediate expressions, the second equation leads to the following parametric equation describing the fastest growing mode:
\begin{equation}
\displaystyle\left(\frac{\rho_0}{\rho_c}\right)^2 \frac{k_R^2}{k^2} =
\frac{c_0^2}{\kappa^2} k^2 \left[ \left(\frac{c_0^2}{\omega_0^2} k^2 - \frac{\rho_0}{\rho_c}\right)^2+ \frac{\rho_0}{\rho_c}\frac{\kappa^2}{\omega_0^2}\right].
\end{equation}

\subsection{Intersection with the line $k_z = k_R$}\label{appendix:fastest2}

\noindent In order to characterize the shape of the perturbations, it would be of interest to determine the intersection of the curve describing the fastest growing mode with the line $k_z = k_R$, which corresponds to spherical perturbations. At the intersection, \mbox{$k^2 = 2 k_R^2$} and $k_R$ satisfies
\begin{equation}
\displaystyle \frac{1}{2}\left(\frac{\rho_0}{\rho_c}\right)^2  =
\frac{c_0^2}{\kappa^2} 2 k_R^2 \left[ \left(\frac{c_0^2}{\omega_0^2} 2 k_R^2 - \frac{\rho_0}{\rho_c}\right)^2+ \frac{\rho_0}{\rho_c}\frac{\kappa^2}{\omega_0^2}\right].
\end{equation}
It reduces to the following polynomial expression, which can be solved numerically:
\begin{equation}
\displaystyle \left(\frac{k_R}{k_0}\right)^6-\frac{\rho_0}{\rho_c} \left(\frac{k_R}{k_0}\right)^4 + \frac{1}{4} \frac{\rho_0}{\rho_c} \left(\frac{\rho_0}{\rho_c}+\frac{\kappa^2}{\omega_0^2}\right) \left(\frac{k_R}{k_0}\right)^2 - \frac{1}{16} \left(\frac{\rho_0}{\rho_c}\right)^2 \frac{\kappa^2}{ \omega_0^2} =0~
\end{equation}
where $k_0 = \omega_0/c_0$ is a characteristic wavenumber depending on the position $R$ through $c_0$.
\end{appendix}

\end{document}